\documentclass{theobservatory}
\usepackage{color,graphicx,rotating}
\usepackage[]{cite}
\usepackage{gensymb}
\usepackage{setspace}
\newcommand{\Msun}{M$_\odot$}

\makeatletter
\renewcommand\@biblabel[1]{(#1)}
\makeatother
\begin{document}

\title{\large SPECTROSCOPIC ORBITS OF THREE DWARF BARIUM STARS}
\author{\it By P. L. North$^1$,  A. Jorissen$^2$,  A. Escorza$^{2,3}$, B. Miszalski$^{4,5}$, and J. Mikołajewska$^6$
\medskip\\
\it $^1$Institute of Physics, Laboratory of Astrophysics,\\
\it Ecole Polytechnique F\'ed\'erale de Lausanne (EPFL), Switzerland\\
\it $^2$Institut d'Astronomie et d'Astrophysique, Universit\'e Libre de Bruxelles, Belgium\\
\it $^3$Instituut voor Sterrenkunde, KU Leuven, Belgium\\
\it $^4$South African Astronomical Observatory \\
\it $^5$Southern African Large Telescope Foundation \\
\it $^6$N. Copernicus Astronomical Center, Polish Academy of Sciences, Warsaw, Poland}
\date{}
\maketitle

{\large Barium stars are thought to result from binary evolution in systems wide enough
to allow the more massive component to reach the asymptotic giant branch and eventually
become a CO white dwarf. While Ba stars were initially known only among
giant or subgiant stars, some were subsequently discovered also on the main
sequence (and known as dwarf Ba stars). We provide here the orbital parameters of three dwarf Ba
stars, completing the sample of 27 orbits published recently by Escorza {\sl et al.} with these three southern targets.
We show that these new orbital parameters are consistent with those of other dwarf Ba stars.
}

\section*{\it\normalsize Introduction}
\label{Sect:intro}
\mbox{}\indent Barium stars are not evolved enough to synthesize in their interiors and dredge up to their surface the s-process
elements (including Ba) that are very abundant  in their atmospheres. On the other hand, we have convincing statistical indications
that they
all belong to SB1-type binary systems \cite{McClure-1990} \cite{Jorissen-2019}, and UV observations have revealed the white dwarf
nature of the companion of some of them\cite{Bohm-Vitense-1984}\cite{Bohm-Vitense-2000}
\cite{Gray-2011}. This leads to the following scenario -- that has become common wisdom -- to explain
their overabundance of s-process elements: the initially more massive component evolved
into an asymptotic giant branch (AGB) star, synthesized and dredged up carbon and s-process elements, expelled its
envelope, part of which was accreted by the "innocent bystander", namely the initially secondary
component, and the latter thereby became the Ba star we observe today while the other
component cooled to a white dwarf. The mass ratio underwent a reversal in the process.
The mass transfer may occur through wind, (wind) Roche lobe overthrow or common envelope evolution \cite{Saladino-2019};
the details are still debated, but the net result is a long orbital period (a few hundred
days to several decades) and a small eccentricity\cite{Jorissen-2019} \cite{Escorza-2019}.

Interestingly, Ba stars were known initially only among G- and K-type giants (they were
dubbed Ba giants when belonging to Population I, and CH giants when belonging to
Population II)\cite{Bidelman-1951}\cite{Keenan-1942} and subgiants (dubbed CH
subgiants)\cite{Bond-1974}, though some of the latter are actually
dwarfs\cite{Escorza-2017}. The existence of main sequence Ba stars was fully recognized
only in the early 1990's\cite{North_Duquennoy-1991}\cite{North_Lanz-1991}\cite{Jorissen_Boffin-1992}
\cite{North-1992}\cite{North-1994a}\cite{North-1994b}\cite{North-2000}, among F- and G-type stars.
The question then arose as to whether giant Ba stars might be the descendants of
Ba dwarfs. Even though the latter appear on average less massive than the former \cite{Escorza-2019},
this is the result of a selection bias (dwarf Ba stars are difficult to detect among the more massive main-sequence A and late-B stars).
From a dynamical viewpoint, the two families appear similar, as revealed  by their almost perfect overlap\footnote{The three
dwarf Ba stars with $P > 2000$~d seemingly falling in the "low-eccentricity gap" ($P > 1000$~d, $e \le 0.05$) of the eccentricity -- period diagram
(Fig.~\ref{Fig:eP}) have error bars on the eccentricity compatible with them lying just at the boundary of the gap.} in the
eccentricity -- period diagram (Fig.~\ref{Fig:eP}).

Ba dwarfs and subgiants are interesting because their mass determination is easier than
for Ba giants. Therefore if a relative astrometric orbit is accessible in addition to the spectroscopic
one, the mass of the companion can be determined. If only the spectroscopic orbit is known,
but for a large enough sample of Ba systems, a statistical estimate of the companion mass
can be obtained assuming random orbit orientations. This was done by e.g. North {\sl et al.}\cite{North-2000}
and by Escorza {\sl et al.}\cite{Escorza-2019}, and resulted in a companion mass close to $0.6$~\Msun,
as expected for a white dwarf. The first sample consisted of 14 orbits, while the second
sample included 27 orbits.

In this work, we present the orbital parameters of three more Ba stars, namely HD~202400, HD~222349,
and HD~224621. The elemental abundances of the first two were determined by North {\sl et
al.}\cite{North-1994a}\cite{North-1994b} and by Luck \& Bond\cite{Luck_Bond-1991} for the third. All three can
be considered as belonging to the main sequence, even though HD~224621 had been classified
as a "CH subgiant", because their surface gravities determined through high-resolution spectroscopy
are $\log g>3.7$.

\section*{\it\normalsize Sample and observations}
\label{Sect:Obs}

\mbox{}\indent The stars HD~202400 and HD~222349 had been classified as F2~Ba~1 and G2~Ba~1 respectively
by L\"{u} {\sl et al.}\cite{Lu-1983} and were studied with high-dispersion spectroscopy by North {\sl et al.}\cite{North-1994a}.
HD~224621 was first classified as a subgiant CH star by Bond\cite{Bond-1974}, then studied with high-dispersion
spectroscopy by Luck \& Bond\cite{Luck_Bond-1991}. Table~\ref{Tab:star_prop} lists the stellar parameters adopted by
these authors and shows that they indeed belong to the main sequence. The magnitudes, colour indices and {\sl Gaia} DR2
parallaxes are also listed. The spectral classification and photometry were taken from the {\sl Simbad} database.
All three stars populate the southern sky, with declinations $\delta < -36\degree$. The masses listed in Table~\ref{Tab:star_prop}
were derived as in ref. \cite{Escorza-2019}.

\begin{table}[htp]
\caption{{\sl Basic properties of the sample stars. The stellar parameters are those adopted by North et al.\cite{North-1994a}
and by Luck \& Bond\cite{Luck_Bond-1991}, except for the mass  (see text). The $B$ and $V$ magnitudes are taken from the Simbad database,
and the parallaxes from the Gaia DR2.}}
\begin{center}
\begin{tabular}{rrlrrrrrrr}
\multicolumn{2}{c}{Star } &Sp. type       &$V$            & $B-V$           & $\pi$        & $T_\mathrm{eff}$ & $\log g$ & [Fe/H] & $M$\\
HD        &  Other id.      &                       &                  &                  &[mas]           &  [K]               & [cgs]       &[dex]   &  [M$_\odot$]\\ \hline
202400 &HIP 105294   & Ap Sr             &$9.18$      &$0.39$       &$5.92$        &$6200$          &$4.0$      &$-0.7$ &	$0.98$\\
             &                      &                       &$\pm 0.01$&$\pm 0.02$&$\pm 0.07$&$\pm 100$    &$\pm 0.2$&$\pm 0.1$ & 	$\pm 0.08$\\
222349 &SAO 247972 &G5/K0+A/F(Sr)& $9.22$       &$0.48$       & $8.16$     &$6000$         &$3.8$   &$-0.9$ & $0.73$ \\
              &                      &                       &$\pm 0.01$&$\pm 0.02$&$\pm 0.10$&$\pm 100$    &$\pm 0.2$&$\pm 0.1$ & $\pm 0.05$\\
224621 &HIP 118266   &  G0III/IV         & $9.55$      &$0.63$        &$12.60$     &$6000$          &$4.0$      &$-0.4$ & $0.90$\\
            &                      &                       &$\pm 0.02$&$\pm 0.03$&$\pm 0.30$ &$\pm 200$    &$\pm 0.3$&$\pm 0.1$&$\pm 0.06$\\ \hline
\end{tabular}
\end{center}
\label{Tab:star_prop}
\end{table}%

The radial-velocity (RV) observations were carried out mainly with the {\sl Coravel} spectrovelocimeter \cite{Baranne-1979}
attached to the 1.54-m Danish telescope at ESO-La Silla (Chile). For HD~202400 and HD~222349, recent observations
were carried out with the 11-m {\sl Southern African Large Telescope} ({\sl SALT} \cite{Buckley-2006}
\cite{O'Donoghue-2006}) using the {\sl  High Resolution Spectrograph} ({\sl HRS}\cite{Bramall-2010}\cite{Bramall-2012}\cite{Crause-2014}) in the medium
resolution (MR) mode, providing resolving powers $R=43\,000$ and
$R=40\,000$ for the blue and red arms respectively.
The basic data products \cite{Crawford-2010} were reduced with the MIDAS pipeline developed by Kniazev {\sl et al.}\cite{Kniazev-2016} which is based
on the ECHELLE \cite{Ballester-1992} and FEROS \cite{Stahl-1999} packages. Heliocentric corrections were applied to the data
using VELSET of the RVSAO package \cite{Kurtz-1998}.
While the {\sl Coravel} radial velocities are obtained at the hardware level
by a physical cross-correlation between the observed spectrum and a mask based on the spectrum of Arcturus, the {\sl HRS}
velocities are obtained {\sl a posteriori} by a digital cross-correlation between the observed spectrum and F0 or G2 masks\cite{Escorza-2019}.
These two stars were also observed once each with the {\sl Coralie} spectrograph attached to the Swiss 1.2-m telescope at
ESO-La Silla, and their radial velocities obtained as well by cross-correlation with an appropriate mask. A few RV measurements
(five for HD 202400, five for HD 222349, and one for HD 224621) are based on spectra taken with the {\sl CES} spectrograph attached
to the 1.4-m {\sl Coudé Auxiliary Telescope} ({\sl CAT}) at ESO-La Silla with a resolving power $R=60\,000$. The RV values are
averages over a few lines fitted by Gaussians, and their adopted errors are the rms dispersion of the RVs given by the individual
lines. A single RV estimate for HD 202400 is based on the positions of the Mg I $\lambda5172$, $\lambda 5183$ lines
of the Mg I triplet and of the two lines of the Na I D doublet, measured on a spectrum taken with the {\sl FEROS} instrument
attached to the 1.52-m telescope at ESO-La Silla. The radial velocities are listed in Table~\ref{Tab:star_RV}.
\begin{table}[htp]
\caption{{\sl Measured radial velocities for the sample stars. Columns 3 and 7 indicate the instrument used.}}
\begin{center}
\begin{tabular}{rllrrllrr}
Star       &HJD              & Instr.    or     & RV                  & $\sigma_{RV}$ &HJD              & Instr.            & RV                  & $\sigma_{RV}$\\
HD        &$ -2400000$ &     ref.           &[km\,s$^{-1}$]  &[km\,s$^{-1}$]    &$ -2400000$ &                     &[km\,s$^{-1}$]  &[km\,s$^{-1}$]  \\ \hline
202400 &$47831.533$&  CES            &$-24.53$         &$2.0$                 &$50705.654$&   Coravel      &$-27.64$          &$3.66$\\
             &$47833.579$&  CES            &$-27.72$         &$1.52$               &$51320.925$&   Coralie       &$-15.40$          &$1.0$  \\
             &$47834.572$&  CES            &$-26.94$         &$1.97$               &$51449.636$&   FEROS      &$-16.10$          &$1.44$\\
             &$48765.938$&  Coravel       &$-22.10$         &$2.8$                 &$58026.336$&   HRS          &$-18.38$          &$0.92$\\
             &$49235.658$&  CES            &$-27.09$         &$0.56$               &$58051.258$&   HRS           &$-17.86$          &$0.73$\\
             &$49237.584$&  CES            &$-26.36$         &$1.76$               &$58061.290$&   HRS           &$-18.04$          &$1.23$\\
             &$49522.860$&  Coravel       &$-24.67$         &$2.30$               &$58064.262$&   HRS           &$-17.89$          &$1.11$\\
             &$49526.858$&  Coravel       &$-21.43$         &$2.42$               &$58238.651$&   HRS           &$-12.92$          &$0.67$\\
             &$49613.607$&  Coravel       &$-24.24$         &$3.16$               &$58292.501$&   HRS           &$-12.80$          &$0.74$\\
             &$49880.838$&  Coravel       &$-12.39$         &$1.76$               &$58329.501$&   HRS           &$-12.80$          &$0.75$\\
             &$50083.534$&  Coravel       &$-17.05$         &$2.99$               &$58361.458$&   HRS          &$-14.87$          &$1.66$\\
             &$50274.821$&  Coravel       &$-25.62$         &$2.27$               &$58413.347$&   HRS          &$-16.49$          &$0.60$\\
             &$50410.513$&  Coravel       &$-27.21$         &$2.98$               &                     &                     &                        &           \\ \hline
222349 &$47830.618$&  CES            &$39.28$          &$0.54$               &$50083.528$&  Coravel       &$32.30$          &$0.40$ \\
             &$48842.873$&  Coravel       &$30.36$          &$0.34$               &$50276.941$&  Coravel       &$35.61$          &$0.42$ \\
             &$48887.707$&  CES            &$30.09$          &$0.27$               &$50410.529$&  Coravel       &$36.91$          &$0.39$ \\
             &$49234.652$&  CES            &$27.28$          &$0.41$               &$50705.751$&  Coravel       &$39.88$          &$0.44$ \\
             &$49236.648$&  CES            &$27.43$          &$0.58$               &$51320.935$&  Coralie        &$38.73$          &$0.09$ \\
             &$49237.665$&  CES            &$27.20$          &$0.41$               &$58084.343$&  HRS            &$28.24$          &$0.15$ \\
             &$49522.947$&  Coravel       &$27.55$          &$0.42$               &$58097.304$&  HRS            &$27.86$          &$0.16$ \\
             &$49610.759$&  Coravel       &$28.16$          &$0.43$               &$58355.587$&  HRS            &$25.39$          &$0.12$ \\
             &$49880.923$&  Coravel       &$30.72$          &$0.42$               &$58409.275$&  HRS            &$26.42$          &$0.03$ \\ \hline
224621 &$43054.791$&  LB91           &$14.0  $          &$0.5$                 &$47400.747$&  Coravel       &$7.82$            &$0.36$ \\
             &$45544.942$&  Coravel       &$  9.11$          &$0.40$               &$47514.549$&  Coravel       &$13.20$          &$0.32$ \\
             &$45979.627$&  Coravel       &$ 14.67$         &$0.34$               &$47756.814$&  Coravel       &$6.53$           &$0.35$ \\
             &$46274.842$&  Coravel       &$ 12.22$         &$0.37$               &$47783.732$&  Coravel       &$7.37$           &$0.33$ \\
             &$46338.704$&  Coravel       &$ 21.77$         &$0.34$               &$48055.927$&  Coravel       &$5.78$           &$0.35$ \\
             &$46395.566$&  Coravel       &$ 20.21$         &$0.34$               &$48074.928$&  Coravel       &$6.03$           &$0.35$ \\
             &$46399.535$&  Coravel       &$ 19.12$         &$0.34$               &$48160.791$&  Coravel       &$18.42$         &$0.33$ \\
             &$46632.880$&  Coravel       &$ 20.19$         &$0.36$               &$48463.862$&  Coravel       &$17.00$         &$0.31$ \\
             &$46667.773$&  Coravel       &$ 22.20$         &$0.32$               &$48842.861$&  Coravel       &$21.81$         &$0.32$ \\
             &$46687.751$&  Coravel       &$ 20.85$         &$0.33$               &$48872.783$&  Coravel       &$18.37$         &$0.35$ \\
             &$46692.706$&  Coravel       &$20.97 $         &$0.39$               &$49235.747$&  CES            &$10.26$         &$0.38$ \\
             &$46724.637$&  Coravel       &$17.06 $         &$0.32$               &$49522.898$&  Coravel       &$13.11$         &$0.36$ \\
             &$46728.617$&  Coravel       &$ 16.26$         &$0.30$               &$49610.765$&  Coravel       &$5.87$           &$0.34$ \\
             &$47043.744$&  Coravel       &$ 15.64$         &$0.33$               &$49880.926$&  Coravel       &$6.97$           &$0.40$ \\
             &$47049.782$&  Coravel       &$ 14.08$         &$0.35$               &$50083.546$&  Coravel       &$19.95$         &$0.37$ \\
             &$47069.667$&  Coravel       &$ 11.30$         &$0.35$               &$50276.937$&  Coravel       &$11.84$          &$0.36$ \\
             &$47100.598$&  Coravel       &$ 7.17  $         &$0.33$               &$50410.535$&  Coravel       &$18.91$         &$0.36$ \\
             &$47104.615$&  Coravel       &$ 7.20  $         &$0.35$               &$50706.750$&  Coravel       &$21.01$         &$0.39$ \\
             &$47370.933$&  Coravel       &$12.41$          &$0.38$               &$50795.536$&  Coravel       &$6.66$           &$0.49$ \\ \hline
\end{tabular}
\end{center}
\label{Tab:star_RV}
\end{table}%

\section*{\it\normalsize Results}
\label{Sect:results}

\mbox{}\indent The search for the most probable orbital period and the determination of the orbital parameters were made
using the DACE software, that is publicly available (\verb+https://dace.unige.ch+) and is optimized for the search of exoplanets
but also appropriate for the analysis of binary stars. The analysis is made in three steps: the first consists in a periodogram,
on which the user can select the most probable period; the second is a first keplerian fit of the orbital parameters, following
the method described by Delisle {\sl et al.}\cite{Delisle-2016}; the third and final step consists in a Monte-Carlo Markov Chain
(MCMC) that determines final values and realistic errors for the orbital parameters, as described by Dìaz {\sl et al.}\cite{Diaz-2014},
\cite{Diaz-2016}. The MCMC even allows us to adjust possible RV zero-point offsets between different instruments. We have
determined the orbital parameters both using this possibility and under the assumption of negligible offsets; in general the
former option remains unsatisfactory in that it artificially lowers the reduced chi-square, so we rather adopted the latter option.
The resulting orbital parameters are displayed in Table~\ref{Tab:orb_par}.
The RV curves are shown in Figs.~\ref{Fig:rv_202400}, \ref{Fig:rv_222349}, and \ref{Fig:rv_224621}.
\begin{table}[htp]
\caption{{\sl Orbital parameters of the sample stars obtained using the DACE code, including the MCMC.
The errors quoted correspond to $1~\sigma$, i.e. to the $68.27$\% confidence interval.}}
\begin{center}
\begin{tabular}{rrrrrrrrrrr}
Star           &$P_\mathrm{orb}$& $e$     &$T_\mathrm{p}$ [HJD&$\omega$&$K_1$           &$\gamma$     &f(m)                 &$a_1\sin i$   &$\sigma_\mathrm{res}$& N \\
HD              & [d]                       &                     & -2400000]        &[\degree]  &[km\,s$^{-1}$]&[km\,s$^{-1}$]&[M$_\odot$]    &[AU]             &[km\,s$^{-1}$] &      \\ \hline
202400       & $1396.6$            & $0.222$       &$48523$           &$14$         &$7.26$          &$-21.56$        &$0.0513$        &$0.906$       &$0.89$            &$25$ \\
                  & $\pm7.3$            & $\pm.067$  &$\pm133$          &$\pm33$   &$\pm.46$      &$\pm.42$       &$\pm.0100$    &$\pm.055$   &                        &         \\
222349       & $3018.6$            & $0.112$      &$54964$         &$123.9$     &$6.89$          &$33.657$       &$0.1004$         &$1.899$        &$0.268$          &$18$ \\
                   & $\pm5.6$            & $\pm.023$  &$\pm45$       &$\pm5.7$&$\pm.11$         &$\pm.076$   &$\pm.0048$     &$\pm.033$        &                       &         \\
224621       & $308.092$            & $0.023$      &$48172.6$       &$316.8$    &$8.173$       &$13.822$       &$0.01741$      &$0.2314$      &$0.376$          &$38$ \\
                   & $\pm.094$            & $\pm.011$  &$\pm15.5$       &$\pm18.2$&$\pm.092$   &$\pm.060$     &$\pm.00059$  &$\pm.0026$  &                       &         \\ \hline
\end{tabular}
\end{center}
\label{Tab:orb_par}
\end{table}

\section*{\it\normalsize Discussion}
\label{Sect:discussion}

\mbox{}\indent Adding these three new orbits to the $27$ 
 published by Escorza {\sl et al.}\cite{Escorza-2019}, 
 the number of orbits for systems hosting dwarf or subgiant Ba stars now amounts to 30.
However, one of the systems studied by Escorza {\sl et al.} is an SB2 system (HD~114520) and could actually be
a triple system, if one accepts the prevailing paradigm about the formation of Ba stars as a given; the WD would then follow a much wider orbit, the period of which remains to be determined.
Another system, HD~48565, is clearly triple but,
as it is an SB1, it is not possible to know whether the WD belongs to the inner or to the outer system. Thus we are left with $28$ binary systems whose statistical properties may be investigated.

\begin{figure}
\resizebox{\hsize}{!}{\includegraphics{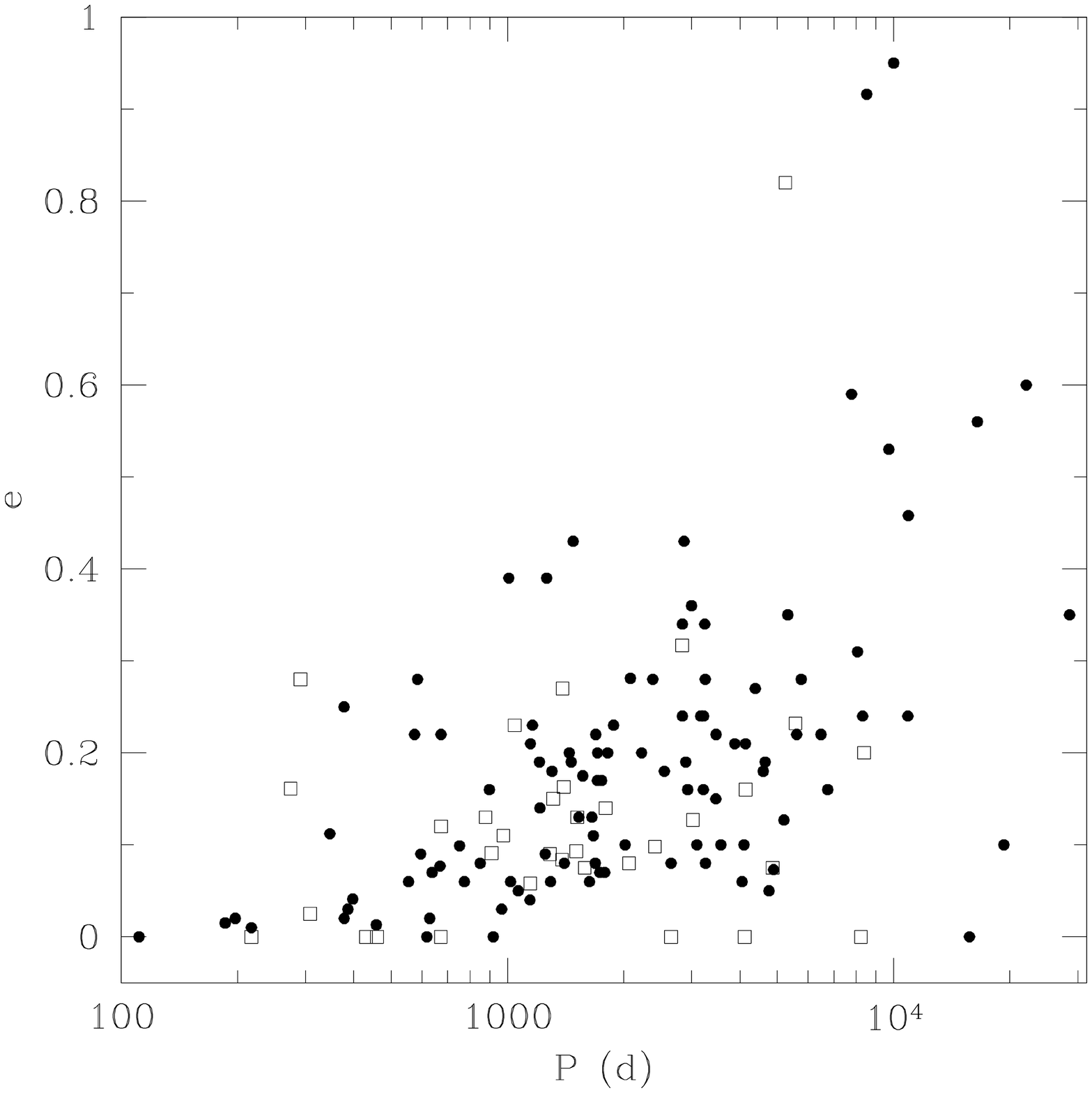}}
\caption[]{
The $e-P$ diagram for dwarf Ba and subgiant CH stars (open squares; from ref. \cite{Escorza-2019} and the present work) and giant Ba and S stars (filled circles;\cite{Jorissen-2019}).
}
\label{Fig:eP}
\end{figure}

\begin{figure}
\resizebox{\hsize}{!}{\includegraphics{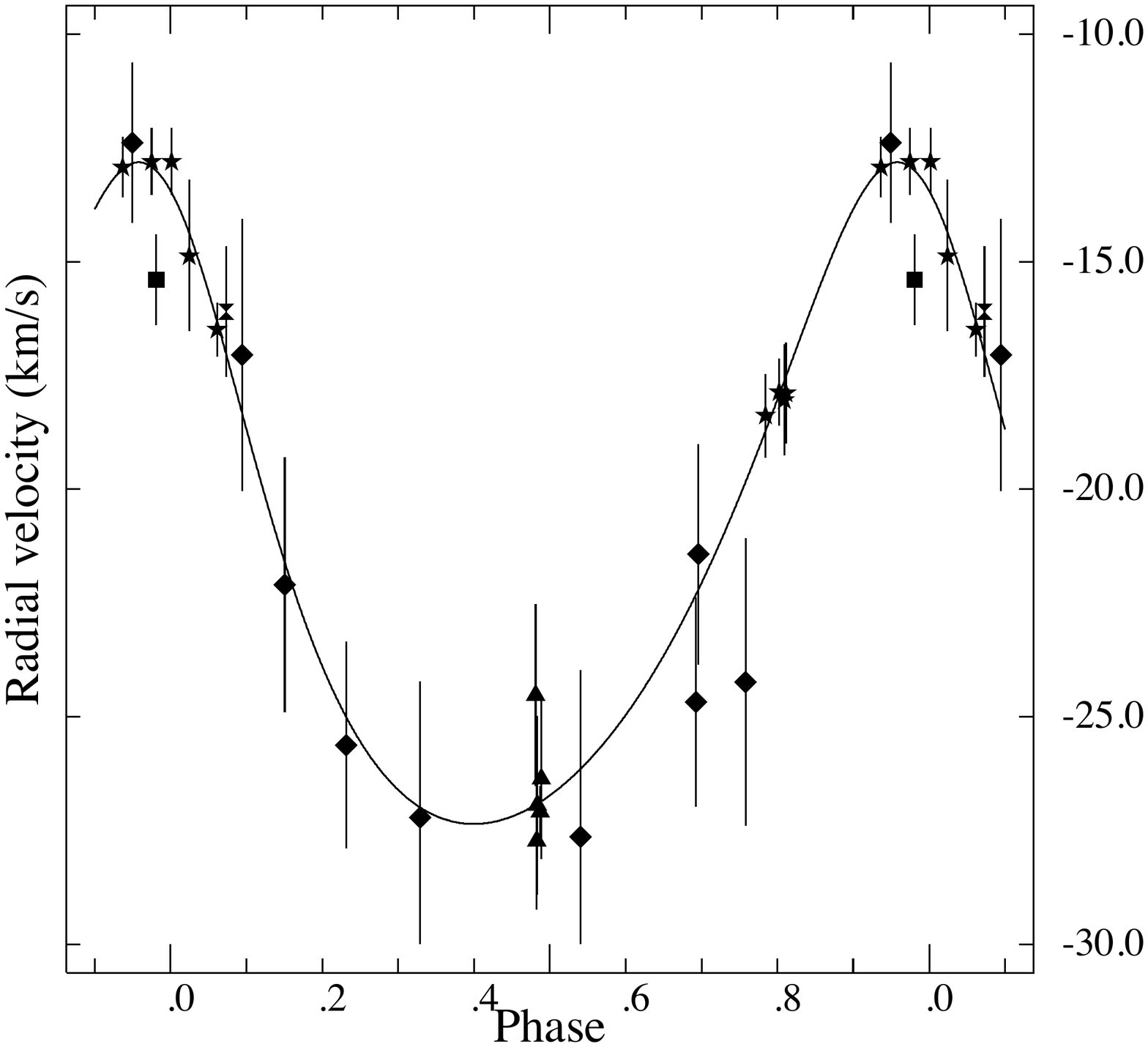}}
\vspace{-5cm}
\caption[]{
Radial velocities (RV) of  HD~202400 as a function of orbital phase. Symbols are as follows: triangles ({\sl CAT}), hourglass ({\sl FEROS}), stars ({\sl SALT}), square ({\sl Coralie}), lozenges ({\sl Coravel}).
The large RV errors and dispersion around the fitted curve are due to a relatively
high projected rotational velocity.
\label{Fig:rv_202400}
}
\end{figure}

\begin{figure}
\resizebox{\hsize}{!}{\includegraphics{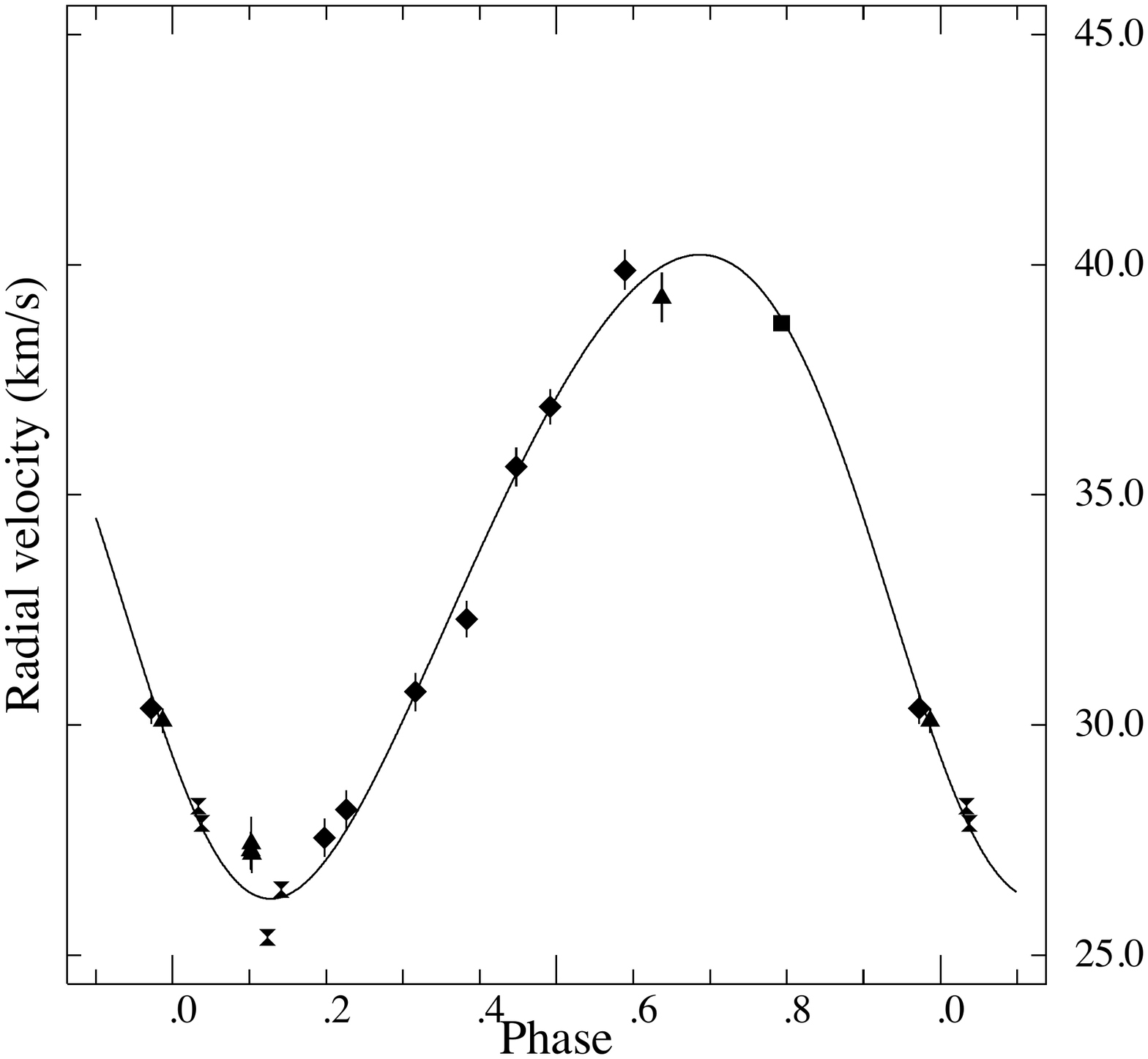}}
\vspace{-5cm}
\caption[]{
Same as Fig.~\ref{Fig:rv_202400}, but for HD 222349.  Symbols are as follows:   triangles ({\sl CAT}), hourglasses ({\sl SALT}), square ({\sl Coralie}), lozenges ({\sl Coravel}).
}
\label{Fig:rv_222349}
\end{figure}

\begin{figure}
\resizebox{\hsize}{!}{\includegraphics{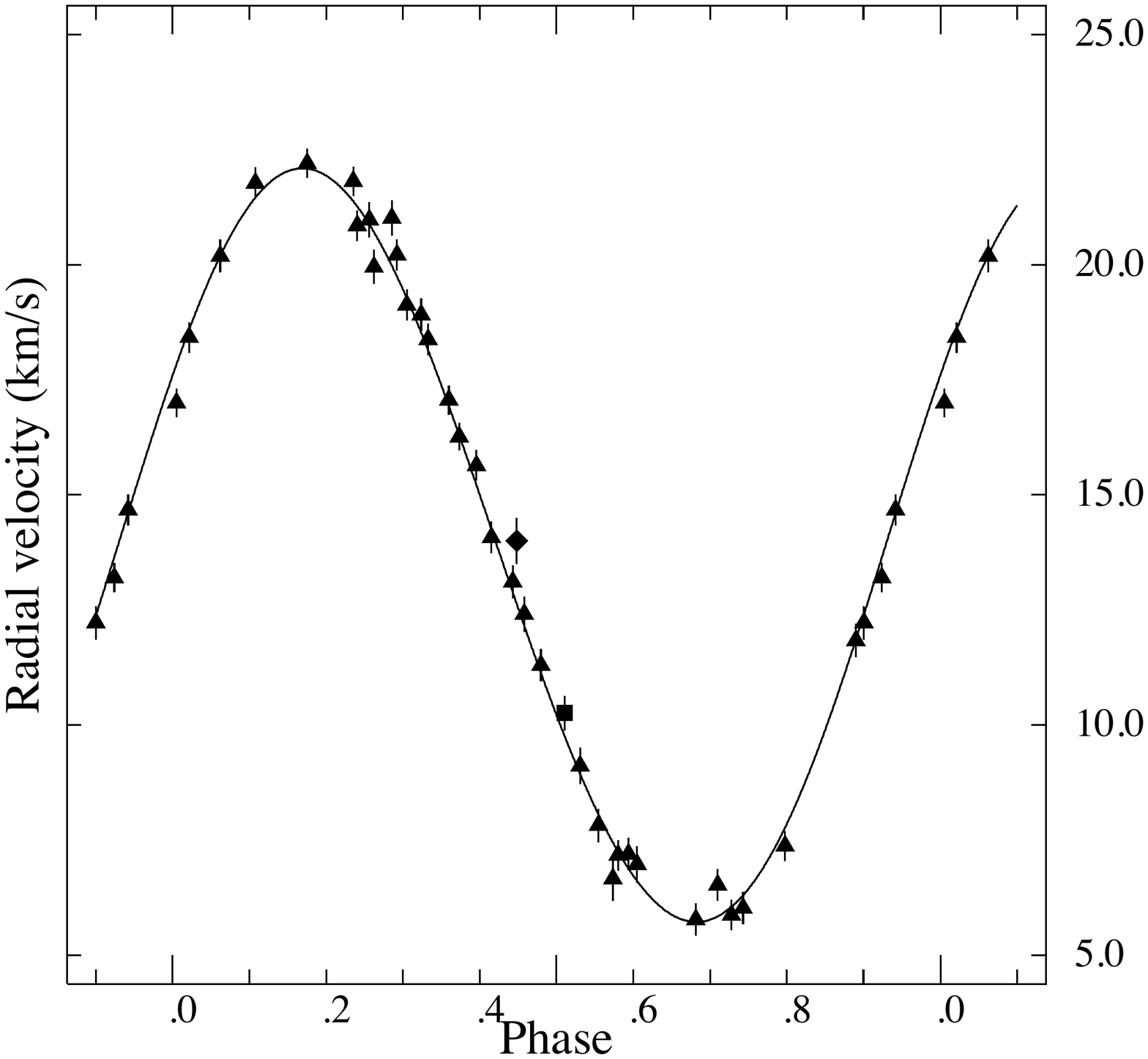}}
\vspace{-5cm}
\caption[]{
Same as Fig.~\ref{Fig:rv_202400}, but for HD 224621. Symbols are as follows:   triangles ({\sl Coravel}), square ({\sl CAT}), lozenge (ref. \cite{Luck_Bond-1991}).
 }
\label{Fig:rv_224621}
\end{figure}

In Fig.~\ref{Fig:eP}, we presented the sample $e - P$ diagram which revealed an almost perfect coincidence between the locations  of the two samples, suggesting  that giant Ba stars
(and their cooler analogues -- the extrinsic S stars) are indeed the descendants of the dwarf Ba stars. Using the BINSTAR evolution code \cite{Davis-2013},
this hypothesis will be further evaluated in a paper in preparation (Escorza {\sl et al.} 2020). 

\begin{figure}
\resizebox{\hsize}{!}{\includegraphics{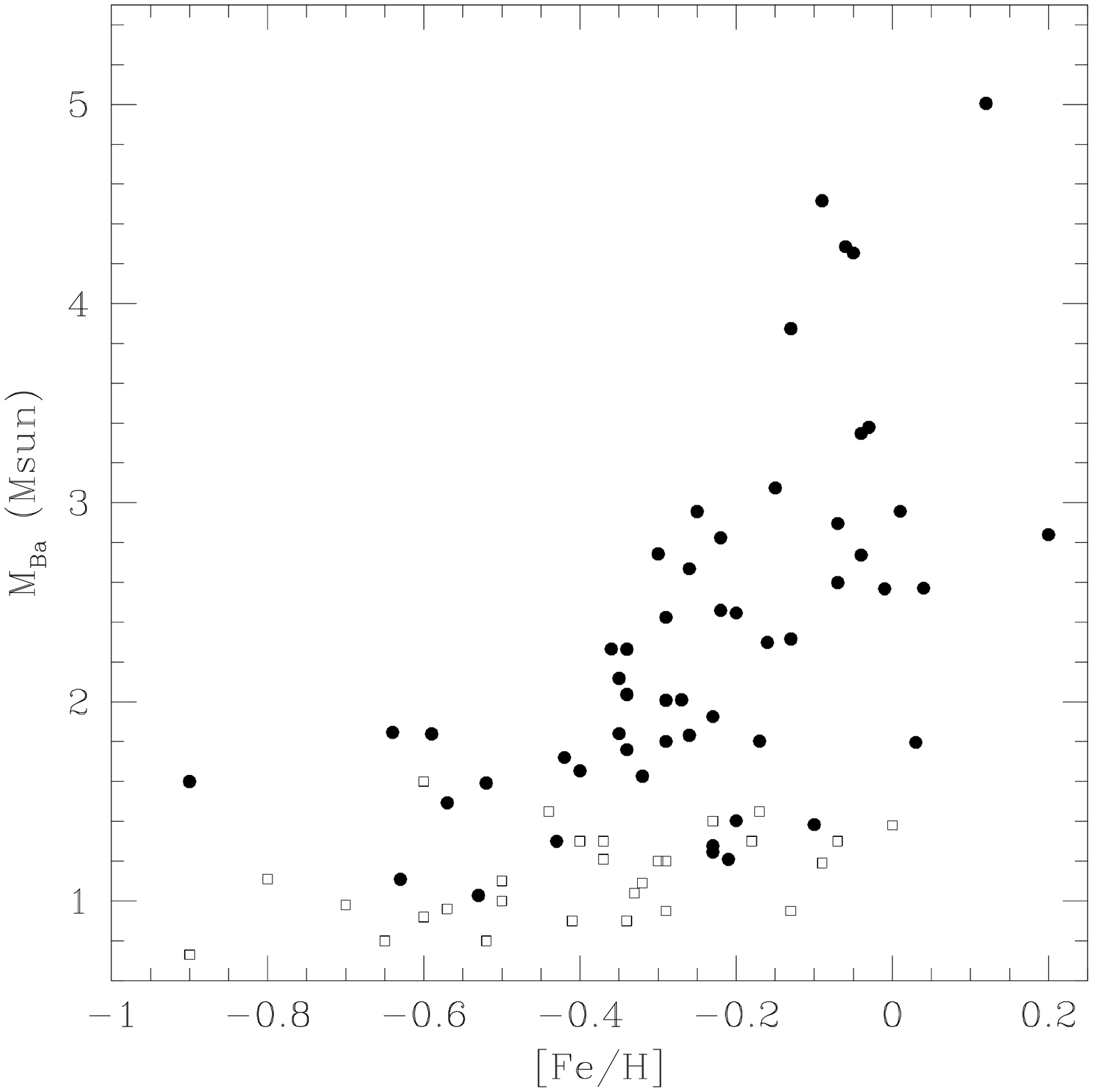}}
\caption[]{
Barium star mass {\sl vs.} [Fe/H], for barium giants (filled circles (ref. \cite{Jorissen-2019})) and barium dwarfs (open squares, from this work and ref. \cite{Escorza-2019}).
}
\label{Fig:mass_metal}
\end{figure}

In Fig.~\ref{Fig:mass_metal} we show the relation between the mass of the Ba star and its metallicity [Fe/H] for both giant (filled circles) and dwarf (open squares) Ba stars.
Here, the triple system HD~48565 is included, because the orbital elements
do not matter, and one can be confident that the non-degenerate, non-Ba component is faint enough that neither the mass nor the metallicity of the primary component
is biassed; the diagram thus includes 29 dwarf Ba stars (open squares). The giant Ba stars are taken from Fig.~17 of Jorissen {\sl et al.}
\cite{Jorissen-2019}. 
For the reason mentioned in the introduction (dwarf Ba stars are difficult to detect among the more massive main-sequence A and late-B stars), dwarf Ba stars so far
appear to be restricted to a narrower mass range than are giant Ba stars. Low-mass Ba stars (with $M < 2$~M$_\odot$) cover a large metallicity range ($-1 \le \mathrm{[Fe/H]} \le 0$), whereas the most massive (giant) Ba stars ($M \ge 4$~M$_\odot$),  being relatively young, are restricted to high metallicities ($-0.2 \le$~[Fe/H]).

It seems appropriate here to dig a bit deeper into the question as to why no dwarf Ba star is known with a mass larger than about $1.6$~M$_\odot$.
A selection bias undoubtedly does occur, due to the fast rotation of most intermediate-mass stars on the main sequence, that
broadens the spectral lines and thus makes abundance determinations difficult. However, some A-type stars are slow rotators,
therefore some of them could in principle be detected as Ba stars. Indeed Ba has been found overabundant in several of these ({\sl e.g.}, by Lemke
\cite{Lemke-1990}, and Takeda {\sl et al.}\cite{Takeda-2008}),
but many of them are classified as Am stars (see, {\sl e.g.}, Fig.~1 of \c{C}ay {\sl et al.}\cite{Cay-2016} for the elemental abundances in two
typical Am stars), and most Am stars are short-period binary systems, with a period distribution that is not compatible with that of Ba stars.
Ba is overabundant in many chemically-peculiar stars,
from those of the HgMn type (among the late B-type stars, see, {\sl e.g.}, Monier et al.\cite{Monier-2016}), to those of the SrCrEu type
(among the mid A-type stars, see, {\sl e.g.}, Guthrie\cite{Guthrie-1969}, Cowley\cite{Cowley-1976}, and Kochukov {\sl et al.}\cite{Kochukhov-2004}).
The problem is that such stars also show underabundance of carbon, contrary to both dwarf and giant Ba stars, and their abundance
anomalies are thought to be due to radiative diffusion rather than mass-transfer in a binary system (see refs. \cite{Michaud-1970}
and \cite{Borsenberger-1984}). Especially telling is the case of the magnetic Ap star HR~3831, where the Ba
overabundance may reach a factor as large as $10^5$ in some places of the stellar surface, according to abundance Doppler imaging
\cite{Kochukhov-2004}: such high overabundances cannot be explained by the mass-transfer scenario invoked for Ba stars.
Furthermore, the high efficiency of radiative diffusion in slowly rotating intermediate-mass stars suggests that any abundance anomaly
due to mass-transfer in a binary system will be quickly superseded and erased by the radiative diffusion mechanism,
making the identification of such systems all the more problematic. Interestingly, the possible link between Am stars and barium stars
had already been proposed by Hakkila\cite{Hakkila-1989}, though his discussion about how to reconcile the discrepant period
distributions remains unconvincing.

Related questions are whether Sirius is a Ba star or not, and why Procyon does not show Ba overabundance. The first question has
been addressed by Landstreet\cite{Landstreet-2011}: Sirius does show a significant ($1.4$~dex) Ba overabundance and it
does have a white dwarf companion, but it
is classified as a mild Am star (A0mA1 Va, according to Gray {\sl et al.}\cite{Gray-2003}) and indeed has abundances typical of Am stars. Its
orbit has a long period ($50.13$~yr) and a rather large eccentricity ($0.591$), but such figures are still within the orbital parameters
covered by Ba stars (see Fig.~\ref{Fig:eP}). For instance, HD~119185\cite{Jorissen-2019} has $P_\mathrm{orb}=60$~yr and $e=0.6$.
It is highly probable that radiative diffusion has been and is still being at work in Sirius A,
but precisely because of this, it remains impossible to know how much processed material it has accreted from its
former AGB companion. Sirius A has a mass $M=2.06$~M$_\odot$, so it is perfectly representative of those stars that are massive enough
to have almost no convective zone, thus permitting radiative diffusion to take place.
On the other hand, the WD has a mass $M=1.02$~M$_\odot$, corresponding to an initial stellar mass $M\sim 5$~M$_\odot$ according
to the semi-empirical initial-final mass relation of Cummings {\sl et al.}\cite{Cummings-2019}. Such high-mass AGB stars with solar
metallicity are unable to yield substantial quantities of s-process elements\cite{Karakas-2016}, suggesting that
radiative diffusion alone is responsible for the Ba overabundance of Sirius A.
Procyon A has a mass $M=1.5$~M$_\odot$
that is similar to that of dwarf Ba stars. It also has a white dwarf companion, with $M=0.60$~M$_\odot$, on an orbit with $P_\mathrm{orb}=40.82$~yr and $e=0.407$,
quite compatible with some orbits of Ba stars. However, its Ba abundance, although slightly enhanced, remains solar within $1\,\sigma$ (and
other heavy s-process elements like Nd and Sm are $2\,\sigma$ below solar) according to Kato \& Sadakane\cite{Kato-1982}.
Why, then, is Procyon A not a Ba star? Although this seems to contradict the mass-transfer scenario at first sight,  at least two
factors may reconcile this scenario with the lack of Ba overabundance: ({\sl i}) the anti-correlation between the overabundance of s-process elements
and the orbital period, and ({\sl ii}) the anti-correlation between that overabundance and [Fe/H] (see ref. \cite{Jorissen-2019}),
and the solar metallicity of Procyon.

\mbox{}\indent 
{\it Acknowledgements}\\
\mbox{}\indent
PN thanks Dr. Maxime Marmier for digging into the old {\sl Coravel database}, thereby providing us
with precious RV values obtained with the southern {\sl Coravel} scanner attached to the 1.54-m Danish
telescope at ESO La Silla, Chile. PN also thanks the colleagues who contributed to this observational
effort, especially Mr. Bernard Pernier of Geneva Observatory and Dr. Jean-Claude Mermilliod of
Institut d'astronomie de l'Université de Lausanne. Some RVs were obtained using the {\sl Coralie} spectrograph attached to
the Swiss 1.2-m {\sl Euler} telescope at ESO La Silla Observatory, Chile, while others were obtained using the
{\sl CES} spectrograph attached to the {\sl CAT} 1.4-m telescope at ESO La Silla. One measurement was made with the
{\sl FEROS} spectrograph attached to the 1.5-m ESO telescope at ESO La Silla.
Recent measurements were gathered with the {\sl Southern African Large Telescope} ({\sl SALT}) using the
{\sl HRS} within the {\sl SALT} programme 2017-1-MLT-010 (PI. B. Miszalski).
This research has been partly funded by the National Science Centre,
Poland, through grant OPUS 2017/27/B/ST9/01940 to JM.
Polish participation in {\sl SALT} is funded by grant No. MNiSW
DIR/WK/2016/07.
Keplerian model initial conditions are computed using the formalism described in Delisle {\sl et al.}\cite{Delisle-2016},
while the MCMC algorithm is described in Díaz {\sl et al.}\cite{Diaz-2014}, \cite{Diaz-2016}. The orbital parameters and their errors were obtained
using the DACE code. PN also thanks Dr. Damien Segransan for his help in using the DACE software.
This research has been partly funded by the Belgian Science Policy Office under contract BR/143/A2/STARLAB.
AE acknowledges support from the Fonds voor Wetenschappelijk Onderzoek Vlaanderen (FWO) under contract ZKD1501-00-W01.
BM acknowledges support from the National Research Foundation (NRF) of South Africa.
This research has made use of the {\sl Simbad} database, operated at CDS, Strasbourg, France.

\bibliographystyle{unsrt}

\end{document}